# Deepfake Detection with Deep Learning: Convolutional Neural Networks versus Transformers

Vrizlynn L. L. Thing
Singapore Technologies Engineering
vriz@ieee.org

*Abstract* — The rapid evolvement of deepfake creation technologies is seriously threating media information trustworthiness. The consequences impacting targeted individuals and institutions can be dire. In this work, we study the evolutions of deep learning architectures, particularly CNNs and Transformers. We identified eight promising deep learning architectures, designed and developed our deepfake detection models and conducted experiments over well-established deepfake datasets. These datasets included the latest second and third generation deepfake datasets. We evaluated the effectiveness of our developed single model detectors in deepfake detection and cross datasets evaluations. We achieved 88.74%, 99.53%, 97.68%, 99.73% and 92.02% accuracy and 99.95%, 100%, 99.88%, 99.99% and 97.61% AUC, in the detection of FF++ 2020, Google DFD, Celeb-DF, Deeper Forensics and DFDC deepfakes, respectively. We also identified and showed the unique strengths of CNNs and Transformers models and analysed the observed relationships among the different deepfake datasets, to aid future developments in this area.

## I. INTRODUCTION

The emergence of deepfake technologies has brought about advancement in the creation of arts [1] and visual effects in films [2] [3] [4]. At the same time, adversaries are abusing deepfakes for the widespread generation and circulation of misinformation. It is a known fact that digital imagery has a powerful effect on human beings [5]. As such, the ease of generating convincing and manipulative deepfakes is seriously threatening the trustworthiness of information. As these deepfakes targeted at individuals and institutions are made widely and readily available on social media platforms, they can lead to serious political, social, financial and legal consequences [6].

To aid in the research of deepfake detection technologies, several datasets namely UADFV [7], DF-TIMIT [8], FF++ [9], Google DFD [10], DFDC Preview [11], Celeb-DF [12], Deeper Forensics [13] and DFDC [14] have been created using various deepfake generation technologies such as FakeApp [15], Faceswap-GAN [16], Deepfakes-Faceswap [17], Face2Face [18], FaceSwap [19], Neural Textures [20], MM/NN Face Swap [21], Neural Talking Heads [22], FSGAN [23], StyleGAN [24]. Based on the release date, synthesis algorithms, quantity and quality of the generated deepfakes, [7] [8] [9] have been categorized as the first generation deepfake datasets. [10] [11] [12] as the second-generation datasets, while [13] [14] were the third-generation datasets in the research community. Table 1 shows the deepfake datasets with their corresponding statistics, generation methods, and release date. These datasets have helped enable the development and evaluation of machine learning, and in particular, deep learning models for enhanced deepfake detection.

| Dataset | Number of real videos | Number of Deepfake videos | Methods | Release Date |
|---|---|---|---|---|
| UADFV [7] | 49 | 49 | [15] | 2018.11 |
| DF-TIMIT [8] | 320 | 320 HQ, 320 LQ | [16] | 2018.12 |
| FF++ [9] | 1000 | 4000 | [17] [18] [19] [20] | 2019.01 |
| Google DFD [10] | 363 | 3068 | Not disclosed | 2019.09 |
| DFDC Preview [11] | 1131 | 4113 | Not disclosed | 2019.10 |
| Celeb-DF [12] | 590 | 5639 | [12] | 2019.11 |
| Deeper Forensics [13] | 48475 | 11000 | [13] | 2020.05 |
| DFDC [14] | 23654 | 104500 | [14] [21] [22] [23] [24] | 2020.06 |

*Table 1: Deepfake Datasets*

In the past few years, deep convolutional neural networks (CNN) have shown their efficacy in continuously pushing the boundaries for better deepfake detection [25] [9] [26] [27] [28] [29] [30] [14]. In the 2019-2020 Kaggle Deepfake Detection Challenge [14], the top 3 submissions were based on the EfficientNet B7 [31] and XceptionNet [32] architectures. However, these prior works were often carried out with self-generated and/or different public datasets, lack in rationalizing the choice of the deep learning architecture and presented with either the overall average AUC or accuracy performance metric, which is often not a suitable metric to be used due the severe data imbalance in the datasets.

In this work, we aim to study the recent evolution of the deep learning architectures, specifically CNN and Transformers. We will design and implement detection models based on the identified promising architectures and explore their specific contributions to Deepfake detection through a comprehensive experimental evaluation over the latest second and third generations of public Deepfake datasets.

This paper is structured as follow. In Section II, we analyse and discuss related works that utilized deep learning models for Deepfake detection. We also shared our key observations in the related works that motivate and shape our work in this paper. In Section III and IV, we discussed the evolution of CNNs and Transformers, respectively. In Section V, we present our proposed work and experiments, and discuss our results. We conclude the paper in Section VI.

## II. RELATED WORK ANALYSIS

In [25], the authors created four CNN models based on the VGG16 [33], ResNet50, ResNet101 and ResNet152 [34] CNN architectures. The authors used self-generated dataset for their training. Evaluation was carried out on the UADFV and DF-TIMIT datasets. They showed that the AUC detection performance achieved by the generated VGG16, ResNet50, ResNet101 and ResNet152 models were 84.5%, 98.7%, 99.1%, and 97.8%, respectively for the UADFV data, 84.6%, 99.9%, 97.6%, 99.4%, respectively for the DF-TIMIT LQ data, and 57.4%, 93.2%, 86.9% and 91.2%, respectively for the DF-TIMIT HQ data. They also trained the Meso-4 and MesoInception-4 [35] models and tested on the two datasets. However, the trained ResNet50 models were shown to outperform the other models for both datasets.

In [9], the authors created CNN models based on the XceptionNet [32] architecture and the FF++ dataset. The trained models achieved detection accuracies of 99.26%, 95.73% and 81% for the FF++ raw, FF++ HQ and FF++ LQ test data, respectively.

In [26], the authors utilized the VGG-Face [36] with ResNet50 architecture for its detection model creation. They monitored the neuron behaviors of deep face recognition to detect the fake faces. Evaluation on the FF++, DFDC and Celeb-DF dataset shows AUC performance of 98.5%, 68% and 66.8%, respectively.

In [27], the authors addressed the detection problem as a multiple instance learning framework. They treated the faces and video as instances and bag, respectively, and built direct mapping from the instance embeddings to the bag prediction. They utilized the XceptionNet architecture to create their detection models. Evaluation on the FF++ raw, FF++ HQ, FF++ LQ, DFDC Preview and Celeb-DF dataset showed detection accuracies of 99.82%, 98.39%, 92.76%, 85.11% and 98.84%, respectively.

In [28], the authors explored two different methods. The first method was based on selecting the entire face as input to the detection. The second method was based on the selection of specific facial regions as inputs. They utilized the XceptionNet architecture to create their detection models. Evaluation of the face input trained models on the UADFV, FF++, DFDC Preview and Celeb-DF dataset showed AUC performance of 100%, 99.4%, 91.17% and 83.6%, respectively. Evaluation results of the eyes, nose, mouth and rest of face input trained models on the UADFV, FF++, DFDC Preview and Celeb-DF dataset are shown in Table 2. The face input trained models outperformed the specific facial region ones across all the four datasets.

In [29], the authors leveraged on blending boundaries in forged face images and adopted the HRNet [37] architecture, to create the detection models. Training was carried out on the FF++ dataset, and tested on the FF++, Google DFD, DFDC Preview and Celeb-DF datasets, and the AUC performance was 98.52%, 95.4%, 80.92% and 80.58%, respectively.

In [30], the authors exploited the source feature inconsistency within forged images to aid in their detection of deepfakes. They adopted the ResNet34 [34] architecture to generate their detection models. Evaluation on FF++, DFDC Preview and Celeb-DF datasets showed AUC performance of 99.79%, 94.38% and 99.98%, respectively. Cross-dataset evaluation was also carried out. Training was carried out using only the real videos in the FF++ dataset, and tested on the FF++, DFD, DFDC Preview, Celeb-DF, Deeper Forensics and DFDC datasets, and the AUC performance was 99.11%, 99.07%, 74.37%, 90.03%, 99.41% and 67.52%, respectively.

|  | Eyes | Nose | Mouth | Rest |
|---|---|---|---|---|
| **UADFV** | 99.7% | 94.7% | 95.4% | 97.3% |
| **FF++** | 92.7% | 86.3% | 93.9% | 85.5% |
| **DFDC Preview** | 83.9% | 81.5% | 79.5% | 76.5% |
| **Celeb-DF** | 77.3% | 64.9% | 65.1% | 60.1% |

*Table 2: AUC Detection Performance of specific facial region based Xception models by Tolosana et al 2021.*

In [14], the authors evaluated the top submissions to the DFDC challenge held on Kaggle. The models were trained and tested on the DFDC dataset. The first submission [38] used an ensemble of 7 detection models created based on the EfficientNet B7 [31] architecture, and achieved an AUC performance of 88.2% and log loss of 0.4279. The second submission [39] utilized the XceptionNet architecture and achieved an AUC performance of 88.3% and log loss of 0.4284. The third model [40] used an ensemble of 3 EfficientNet B7 architectures, and achieved an AUC performance of 88% and log loss of 0.4345.

The evaluation metric commonly used across these prior works is the AUC performance metric. We compiled the AUC evaluation results according to the CNN architecture, training dataset and testing dataset, and presented them in Table 3. Both [9] and [27] utilized the XceptionNet architecture, but were excluded from this table as they used detection accuracy as their performance metric. However, the AUC evaluation results of XceptionNet based models which considered a wider coverage of datasets in a more recent work [28] helped fill this gap. Thus, we have included [28] in the table instead. From Table 3, we made the following key observations.

### A. Key Observations

- It can be challenging utilizing deep CNN architectures to train detection models on the first generation (Generation 1 / Gen-1) datasets. This is especially true for the UADFV and DF-TIMIT datasets due to their limited size. Thus, researchers often resort to generating their own training data, or devise detection solutions based on approaches that are less reliant on the requirement of having a sufficiently large dataset, which is needed for deep learning.

- The Gen-1 datasets have been widely explored in deepfake detection research. They have been used often as the training dataset (especially so for FF++) and/or testing dataset for various prior works, and the relative ease to detect them have also been well-proven, with AUC performance ranging between 98% to 100% across most prior works. Nonetheless, from these prior works, models trained using FF++ and tested on Gen-2 and Gen-3 datasets appear to be relatively weaker in performance.

- The FF++ dataset was expanded in 2020 by including another 1000 deepfakes generated using the

FaceShifter technique [41]. We shall refer to this dataset as FF++ 2020 in this paper. The referred works in this paper had utilized FF++ or its subset.

- The Google DFD dataset was created with deepfake generation technologies that were not disclosed to the public, which could be a reason for it being excluded for consideration in most prior work research. Even with some CNN architecture based prior works (trained using other datasets such as FF++) showing AUC performance exceeding 95%, this dataset may warrant further exploration due to limited exploration.

- The DFDC Preview dataset is an earlier release and subset of the DFDC dataset. Despite that, prior works were having challenges in attempting to achieve a consistently high detection rate when evaluated on the DFDC Preview dataset. During the Kaggle DFDC Challenge, most participants had attempted to create ensembles of detection models based on deep CNN architectures to achieve breakthrough in the detection rates, while trying to stay within the execution testing time limit of 9 hours. A more systematic analysis of CNN models on the DFDC dataset was not explored. Further work to develop and evaluate the efficacies of single models (instead of ensembled models) more thoroughly is needed.

- There is limited well-documented cross-dataset evaluation amongst the Gen-3 datasets. For example, there is limited CNN based models trained using Gen-3 datasets and tested on other datasets, as well as models trained using other datasets and tested on Gen-3 datasets.

Based on these observations, we had included FF++ 2020, Google DFD, Celeb-DF, Deeper Forensics and DFDC datasets for our work in this paper.

| Prior Work | CNN Architecture | Training Dataset | Testing Dataset (AUC Results in %) | | | | | | | |
|---|---|---|---|---|---|---|---|---|---|---|
| | | | Generation 1 (Gen-1) | | | Generation 2 (Gen-2) | | | Generation 3 (Gen-3) | |
| | | | UADFV [7] | DF-TIMIT [8] | FF++ [9] | Google DFD [10] | DFDC Preview [11] | Celeb-DF [12] | Deeper Forensics [13] | DFDC [14] |
| [25] | VGG16 | Self-generated | 84.5 | 84.6 | | | | | | |
| | ResNet50 | Self-generated | 98.7 | 99.9 | | | | | | |
| | ResNet101 | Self-generated | 99.1 | 97.6 | | | | | | |
| | ResNet152 | Self-generated | 97.8 | 99.4 | | | | | | |
| [26] | ResNet50 | FF++, DFDC, Celeb-DF | | | 98.5 | | | | | |
| | ResNet50 | FF++, DFDC, Celeb-DF | | | | | | | | 68 |
| | ResNet50 | Celeb-DF | | | | | | 66.8 | | |
| [28] | XceptionNet | UADFV | 100 | | | | | | | |
| | | FF++ | | | 99.4 | | | | | |
| | | DFDC Preview | | | | | 91.17 | | | |
| | | Celeb-DF | | | | | | 83.6 | | |
| [29] | HRNet | FF++ | | | 98.52 | | | | | |
| | | FF++ | | | | 95.4 | | | | |
| | | FF++ | | | | | 80.92 | | | |
| | | FF++ | | | | | | 80.58 | | |
| [30] | ResNet34 | FF++ | | | 99.79 | | | | | |
| | | DFDC Preview | | | | | 94.38 | | | |
| | | Celeb-DF | | | | | | 99.98 | | |
| | | FF++ (Real) | | | 99.11 | | | | | |
| | | FF++ (Real) | | | | 99.07 | | | | |
| | | FF++ (Real) | | | | | 74.37 | | | |
| | | FF++ (Real) | | | | | | 90.03 | | |
| | | FF++ (Real) | | | | | | | 99.41 | |
| | | FF++ (Real) | | | | | | | | 67.52 |
| [14] | EfficientNet B7 (7-model ensemble) | DFDC | | | | | | | | 88.2 |
| | XceptionNet | DFDC | | | | | | | | 88.3 |
| | EfficientNet B7 (3-model ensemble) | DFDC | | | | | | | | 88 |

*Table 3: Deepfake Detection Results of Prior Works utilizing CNN Architectures*

## III. EVOLUTION OF CONVOLUTIONAL NEURAL NETWORKS

In recent years, CNNs have been demonstrated to exhibit outstanding performance through its powerful learning ability in computer vision, image processing benchmarking competitions, and natural language processing tasks [42]. CNNs leverage on multiple feature extraction stages to automatically learn data representations and have a strong ability to capture signal spatiotemporal dependences. Recent advancements focus on research on different activation and loss functions, parameter optimization, regularizations and most importantly, in CNN's architectural innovations. Significant improvements in the representational capacity of deep CNNs were demonstrated through architectural innovations.

LeNet was the first CNN architecture and was introduced in 1989 [43]. It was a simple CNN which applied back propagation to handwritten zip code recognition. Since then, other CNN architectures such as AlexNet [44] which won the classification and localization tasks at the Large Scale Visual Recognition Challenge 2012 [45] with a deeper CNN model and more channel consideration, InceptionNet [46] that incorporated multi-scale feature extraction and increased the model width with varying sizes of kernels in parallel (instead of only depth increase), VGG (Visual Geometry Group at University of Oxford) [47] which used an architecture with very small (3x3) convolution filters and pushed the depth to 16-19 weight layers showed significant improvements over prior works in 2014, had also been proposed.

In 2016, ResNet [34] was proposed to address the vanishing gradient problem with deeply stacked multi-layers CNNs. Residual connections were introduced to create alternate paths for the gradient to skip the middle layers and reach the initial layers, which allowed extremely deep models with good performance to be trained. In 2017, XceptionNet [32], inspired by InceptionNet, was proposed. In XceptionNet, the Inception modules were replaced by depthwise separable convolutions, and an evaluation on ImageNet showed that XceptionNet was able to achieve better performance over InceptionNet-V3. In 2019, EfficientNet [31] based on the design of balancing the network depth, width and resolution, and a method to uniformly scale all dimensions of depth, width and resolution using a compound coefficient was proposed. A family of eight model architectures, EfficientNet B0 to EfficientNet B7, scaled from the EfficientNet B0 baseline network were created. The authors demonstrated that EfficientNet B7 achieved the highest top-1 accuracy of 84.3% compared to InceptionNet-V4 at 80.0%, XceptionNet at 79%, ResNet152 at 77.8%, and ResNet50 at 76%, on the ImageNet data. In 2020, HRNet [37] was proposed to maintain high-resolution representations to form a stronger backbone for computer vision problems that are position-sensitive. These computer vision problems include object detection, semantic segmentation and human pose estimation. The authors explained that this approach is unlike CNNs such as ResNet and VGG, where the input image was first encoded as a low-resolution representation through a sub-network formed by connecting high-to-low resolution convolutions in series, and the high-resolution representation was recovered from the encoded low-resolution representation. Instead, in HRNet, the high-to-low resolution convolution streams were connected in parallel, and information was exchanged repeatedly across the resolutions. The authors demonstrated that their work delivered top results in semantic segmentation and facial landmark detection, compared to prior works.

## IV. EVOLUTION OF TRANSFORMERS

In 2017, Transformers [48], primarily and initially created for the purpose of performing natural language processing (NLP) tasks, were proposed. However, Transformers saw limited applications to computer vision problems then. For such problems, attention was either applied with CNNs or to replace certain components of CNNs. The overall CNN architecture often remained intact.

In 2020, the vision transformer, VIT [49], was proposed to demonstrate that the direct application of a pure transformer to sequences of image patches, can perform well on image classification tasks. The authors demonstrated that when pre-trained on a large dataset and transferred to smaller image recognition tasks, the VIT model can attain excellent results compared to CNNs such as ResNet. In 2021, the Bidirectional Encoder representation from image Transformers (BEiT) [50], a self-supervised vision representation model was proposed. The authors proposed pretraining the vision transformers with a masked image modeling task, where the original image is first tokenized into visual tokens, randomly masked and fed into the backbone transformer. The pre-training objective was to recover the original visual tokens based on the masked image patches. The model parameters are then fine-tuned on downstream tasks by appending the task layers upon the pre-trained encoder. The authors demonstrated that the BEiT model can achieve 83.2% top-1 accuracy on ImageNet. In the same year, the Swin transformer [51], which built hierarchical feature maps through image patch merging in deeper layers, was proposed. It achieved linear computation complexity to input image size as the self-attention was computed within each local window while allowing cross-window connection. This approach contrasted with previous vision transformers, which produced feature maps of a single low resolution and had quadratic computation complexity to input image size due to global self-attention computation. The authors demonstrated that the Swin model can achieve 84.5% top-1 accuracy, compared to EfficientNet B7 at 84.3%, and VIT at 77.9% on ImageNet. The Class-attention in image Transformers (CaiT) [52] was also proposed to build and optimize deeper transformer networks, specifically for image classification. The authors added a learnable diagonal matrix on the output of each residual block, to allow training of deeper high-capacity image transformers that benefit from depth. They had also separated the transformer layers involving self-attention between patches, from class-attention layers devoted to processed patches' content extraction into a single vector, to avoid contradictory objective of guiding the attention process while processing class embedding. The authors demonstrated that the CaiT model can attain 86.5% top-1 accuracy on ImageNet.

## V. PROPOSED WORK AND EXPERIMENTAL RESULTS

In this work, we proposed the following approach for the development and evaluation of our deepfake detection deep learning models. We first performed a dataset split of each of the five datasets into the respective non-overlapping training, validation and testing data. Except for the DFDC dataset where the training, validation and testing splits were already provided, we split the videos in the other datasets to have at least 10% for validation and 20% for testing, leaving around 70% for the training. We also ensured that specific source videos used to create deepfake videos were kept to within the same split. Thus, this action resulted in the differences in percentage of distributions split across real and fakes while we kept to the minimal percentages as mentioned above. Next, we extracted the frames from the real and deepfake videos, and performed facial detection and extraction, to create balanced real vs fakes in the training and validation datasets. The testing dataset was retained as video data and during the detection process, the video frames and facial regions were extracted, for detections to obtain the test results. The results across the analysed frames were averaged for each video. The splits of the videos in each dataset are shown in Table 4.

We chose ResNet152, XceptionNet, EfficientNet B7 and HRNet as the CNN architectures, and VIT, BEiT, Swin and CaiT as the Transformers, as the underlying architecture for creating our detection models. For each architecture, we implemented the same data pre-processing and training process, and trained each deep learning model for 50 epochs. Finally, we selected the model with the highest validation accuracy under each architecture, for evaluations with the testing dataset. We also performed cross datasets evaluation tests. The balanced accuracy and AUC results are shown in Table 5 and Table 6, respectively.

**Best Detection Results**

We observed that the highest accuracies achieved for detecting FF++ 2020, Google DFD, Celeb-DF, Deeper Forensics and DFDC test deepfakes, were at 88.74%, 99.53%, 97.68%, 99.73% and 92.02%, respectively. The highest AUC achieved for detecting FF++ 2020, Google DFD, Celeb-DF, Deeper Forensics and DFDC test deepfakes, were at 99.95%, 100%, 99.88%, 99.99% and 97.61%, respectively.

**Same Train-to-Test Dataset Evaluations**

Both the CNNs and Transformers models achieved similar good performance (> 85% accuracies, and > 98% AUC) for the FF++ 2020 and Google DFD datasets when tested against their respective test datasets, except for VIT (at around 9 to 12% lower accuracies, and around 1.5 to 7.5% lower AUC). As we moved towards larger and newer generation datasets, the CaiT model dropped in performance when tested against its respective dataset. The CNNs achieved either similar or better performance compared to the rest of the Transformers models in this case. In general, HRNet, XceptionNet and EfficientNet B7 demonstrated their stronger detection performance for their respective test datasets.

**Cross Datasets Evaluations**

The models trained with the FF++ 2020 dataset worked well on the Celeb-DF test dataset. The XceptionNet, VIT, BEiT and CaiT models trained with the F++ 2020 dataset achieved 72.03% to 75.95% accuracies when tested on the Cele-DF dataset. The XceptionNet, EfficientNet B7, VIT, BEiT and CaiT models trained with the F++ 2020 dataset achieved 84.2% to 87.54% AUC when tested on the Cele-DF dataset. Some models trained with the FF++ 2020 dataset worked well with the Google DFD test dataset. The EfficientNet B7 and Swin models trained with the FF++ 2020 dataset achieved 71.1% and 84.24% accuracies when tested on the Google DFD dataset. The XceptionNet, EfficientNet B7, HRNet, BEiT, Swin and CaiT models trained with the FF++ 2020 dataset achieved 91.24% to 94.36% AUC when tested on the Google DFD dataset. On the other hand, The VIT, BeiT and CaiT Transformers models also demonstrated better results on the other test datasets when compared to the CNNs models.

| | | Real videos | Fake videos | Real % | Fake % |
|---|---|---|---|---|---|
| **FF++ 2020** | Training | 700 | 2930 | 70 | 58.6 |
| | Validation | 100 | 760 | 10 | 15.2 |
| | Testing | 200 | 1310 | 20 | 26.2 |
| | Total | 1000 | 5000 | 100 | 100 |
| **Google DFD** | Training | 243 | 1349 | 66.94 | 43.97 |
| | Validation | 43 | 752 | 11.85 | 24.51 |
| | Testing | 77 | 967 | 21.21 | 31.52 |
| | Total | 363 | 3068 | 100 | 100 |
| **Celeb-DF** | Training | 391 | 3554 | 66.27 | 63.03 |
| | Validation | 60 | 740 | 10.17 | 13.12 |
| | Testing | 139 | 1345 | 23.56 | 23.85 |
| | Total | 590 | 5639 | 100 | 100 |
| **Deeper Forensics** | Training | 33509 | 7293 | 69.13 | 66.3 |
| | Validation | 5110 | 1144 | 10.54 | 10.4 |
| | Testing | 9856 | 2563 | 20.33 | 23.3 |
| | Total | 48475 | 11000 | 100 | 100 |
| **DFDC** | Training | 19154 | 100000 | 80.98 | 95.69 |
| | Validation | 2000 | 2000 | 8.46 | 1.91 |
| | Testing | 2500 | 2500 | 10.57 | 2.39 |
| | Total | 23654 | 104500 | 100 | 100 |

*Table 4: Dataset Splits*

The models trained with the Google DFD dataset worked well on the Celeb-DF test dataset. The XceptionNet, VIT, Swin and CaiT models trained with the Google DFD dataset can achieve 73.1% to 80.75% accuracies when tested on the Celeb-DF dataset. The XceptionNet, HRNet, BEiT, Swin and CaiT models achieved 86.84% to 92.61% AUC when tested on the Celeb-DF dataset. When tested on the DFDC dataset, all the models trained on the Google DFD dataset achieved an average accuracy of 65.31%, with the XceptionNet model achieving 71.14% accuracy. The models also achieved an average AUC of 73.53% and 86.72% when tested on the DFDC and Deeper Forensics datasets, respectively.

The models trained with the Celeb-DF dataset worked well on the Google DFD test dataset. The ResNet152, XceptionNet, EfficientNet B7, VIT and BEiT models achieved 71.2% to 75.18% accuracies, and 86.3% to 91.2% AUC, when tested on the Google DFD test dataset.

The models trained on the Deeper Forensics dataset performed badly when tested on the other datasets. One possible reason was that this dataset was very different from the other datasets.

| Train Data | DL Arch | Test Data (Balanced Accuracy Results in %) | | | | |
|---|---|---|---|---|---|---|
| | | FF++ 2020 | Google DFD | Celeb-DF | Deeper Forensics | DFDC |
| FF++ 2020 | ResNet | 87.04 | 58.73 | 67.60 | 48.16 | 55.80 |
| | Xception | 88.32 | 68.31 | 75.06 | 46.82 | 53.82 |
| | Efficient | 87.78 | 71.10 | 66.99 | 49.67 | 55.72 |
| | HRNet | **88.74** | 60.42 | 70.27 | 38.55 | 52.50 |
| | VIT | 75.73 | 52.39 | 74.67 | **75.88** | 68.86 |
| | BEiT | 86.82 | 66.47 | **75.95** | 57.39 | 59.74 |
| | Swin | 87.25 | 51.90 | 66.08 | 33.23 | 58.86 |
| | CaiT | 85.48 | **84.24** | 72.03 | 54.92 | 56.50 |
| Google DFD | ResNet | 51.09 | 97.64 | 64.35 | 51.40 | 63.16 |
| | Xception | 48.97 | 98.99 | **80.75** | 61.73 | **71.14** |
| | Efficient | 51.05 | 99.12 | 60.89 | 51.61 | 63.88 |
| | HRNet | 52.90 | **99.53** | 64.13 | 54.02 | 64.26 |
| | VIT | 52.69 | 87.20 | 73.10 | 67.17 | 64.50 |
| | BEiT | 53.58 | 99.33 | 67.45 | 64.06 | 69.98 |
| | Swin | **59.50** | 96.60 | 79.59 | 62.33 | 62.10 |
| | CaiT | 51.11 | 99.25 | 79.68 | **67.73** | 63.44 |
| Celeb-DF | ResNet | 51.22 | 71.98 | 96.96 | 52.20 | 53.28 |
| | Xception | **51.41** | 75.18 | **97.68** | 53.12 | **54.10** |
| | Efficient | 50.38 | 73.01 | 96.43 | 50.96 | 51.78 |
| | HRNet | 50.23 | 62.77 | 96.47 | 50.35 | 51.86 |
| | VIT | 51.03 | 59.88 | 87.21 | **53.96** | 53.08 |
| | BEiT | 51.34 | 71.20 | 87.51 | 53.12 | 52.76 |
| | Swin | 50.53 | 74.97 | 76.58 | 51.01 | 51.18 |
| | CaiT | 50.19 | 58.27 | 56.17 | 51.25 | 50.90 |
| Deeper Forensics | ResNet | 40.00 | 49.95 | 50.00 | 99.72 | 50.42 |
| | Xception | 40.00 | **50.39** | 50.00 | **99.73** | 50.44 |
| | Efficient | 40.00 | 49.95 | 50.00 | 99.57 | 50.48 |
| | HRNet | 40.00 | 49.95 | 50.00 | 99.70 | 50.44 |
| | VIT | 40.00 | 49.95 | 50.00 | **99.73** | 50.42 |
| | BEiT | 40.00 | 49.95 | 50.00 | **99.73** | **50.52** |
| | Swin | 40.00 | 50.34 | 50.00 | **99.73** | 50.32 |
| | CaiT | 40.00 | 49.95 | 50.00 | **99.73** | 50.22 |
| DFDC | ResNet | 56.60 | 90.51 | 75.12 | 67.91 | 86.84 |
| | Xception | 55.98 | 88.48 | 82.13 | 69.98 | 88.08 |
| | Efficient | 59.17 | 93.01 | **89.27** | 73.85 | **92.02** |
| | HRNet | 56.17 | 87.76 | 80.47 | 68.26 | 87.74 |
| | VIT | 56.27 | 81.42 | 77.63 | 76.31 | 86.62 |
| | BEiT | **59.27** | 95.44 | 77.31 | **79.40** | 89.16 |
| | Swin | 59.00 | **96.20** | 84.34 | 70.26 | 84.50 |
| | CaiT | 52.90 | 74.87 | 55.24 | 53.76 | 67.40 |

Table 5: Accuracies of Our Detection Models

| Train Data | DL Arch | Test Data (AUC Results in %) | | | | |
|---|---|---|---|---|---|---|
| | | FF++ 2020 | Google DFD | Celeb-DF | Deeper Forensics | DFDC |
| FF++ 2020 | ResNet | 99.21 | 87.59 | 78.59 | 42.42 | 68.37 |
| | Xception | 99.70 | 91.24 | 84.20 | 39.44 | 65.18 |
| | Efficient | 99.45 | 92.04 | **87.54** | 46.88 | 68.44 |
| | HRNet | **99.95** | 92.27 | 77.93 | 16.98 | 62.26 |
| | VIT | 92.29 | 84.83 | 84.35 | **86.02** | 76.30 |
| | BEiT | 98.76 | 92.54 | 84.80 | 73.63 | 70.41 |
| | Swin | 98.81 | 92.25 | 72.57 | 18.51 | 65.70 |
| | CaiT | 99.14 | **94.36** | 84.52 | 74.46 | 62.38 |
| Google DFD | ResNet | 60.46 | 99.91 | 79.60 | 85.62 | 69.67 |
| | Xception | 57.37 | 99.90 | 90.55 | 92.89 | **78.48** |
| | Efficient | 63.34 | 99.94 | 86.71 | 75.29 | 73.44 |
| | HRNet | 61.80 | **100.00** | 92.61 | 83.39 | 74.62 |
| | VIT | 62.60 | 98.11 | 81.97 | 88.68 | 72.56 |
| | BEiT | 63.82 | **100.00** | 90.05 | 93.20 | 76.79 |
| | Swin | **67.35** | 99.63 | 86.84 | 78.64 | 71.70 |
| | CaiT | 58.84 | 99.92 | 88.91 | **96.03** | 70.96 |
| Celeb-DF | ResNet | 65.68 | 86.30 | 99.65 | 82.12 | 70.24 |
| | Xception | 65.61 | 89.32 | 99.77 | 85.76 | **70.46** |
| | Efficient | 65.88 | 88.55 | **99.88** | 76.26 | 69.42 |
| | HRNet | 64.12 | 81.76 | 99.87 | 71.13 | 67.25 |
| | VIT | 65.08 | 79.87 | 97.54 | **88.32** | 69.12 |
| | BEiT | 63.41 | 88.09 | 98.37 | 71.34 | 65.85 |
| | Swin | **66.52** | 91.20 | 98.84 | 74.48 | 63.91 |
| | CaiT | 61.66 | 84.14 | 96.76 | 65.45 | 59.42 |
| Deeper Forensics | ResNet | 41.43 | 54.56 | 50.32 | 99.99 | 55.69 |
| | Xception | 40.36 | 49.28 | **55.71** | 99.99 | 53.95 |
| | Efficient | 40.88 | 53.79 | 50.99 | 99.99 | 56.65 |
| | HRNet | 40.71 | **58.18** | 52.76 | 99.99 | 58.60 |
| | VIT | **43.46** | 49.52 | 48.25 | 99.99 | 59.14 |
| | BEiT | 41.30 | 52.50 | 49.38 | 99.99 | **60.74** |
| | Swin | 38.79 | 51.85 | 48.93 | 99.99 | 53.15 |
| | CaiT | 41.94 | 56.94 | 54.19 | 99.99 | 59.85 |
| DFDC | ResNet | 73.34 | 96.94 | 90.21 | 95.33 | 94.81 |
| | Xception | 71.19 | 95.69 | 95.21 | 93.02 | 95.96 |
| | Efficient | **74.22** | 98.42 | **97.14** | 93.74 | **97.61** |
| | HRNet | 69.71 | 94.85 | 95.57 | 90.78 | 94.97 |
| | VIT | 69.39 | 92.51 | 93.35 | **98.36** | 93.41 |
| | BEiT | 73.89 | 99.42 | 96.22 | 97.37 | 95.90 |
| | Swin | 72.02 | **99.76** | 92.20 | 95.60 | 92.62 |
| | CaiT | 70.06 | 95.72 | 89.87 | 95.01 | 85.13 |

Table 6: AUC Results of Our Detection Models

The models trained on the DFDC dataset performed very well on the Google DFD test dataset. The ResNet152, EfficientNet B7, BeiT and Swin models achieved 90.51% to 96.2% accuracies and 96.94% to 99.76% AUC, when tested on the Google DFD test dataset. The eight models achieved an average 88.46% accuracy and 96.66% AUC when tested on the Google DFD dataset. The models, except for CaiT, also performed well on the Celeb-DF test dataset, and can achieve 75.12% to 89.27% accuracies and 90.21% to 97.14% AUC.

The EfficientNet B7, VIT, BeiT and Swin models achieved 70.26% to 79.4% accuracies and 93.74% to 98.36% AUC when tested on the Deeper Forensics test dataset. Thus, models trained with the DFDC dataset were able to perform well across all the other test datasets, except for FF++ 2020.

**Further Analysis**

In general, the HRNet, XceptionNet, EfficientNet B7 CNNs models performed very well consistently when trained and tested on the same dataset. However, HRNet models did not work very well on cross datasets evaluations. On the other hand, the XceptionNet, and the VIT, BeiT and Swin Transformers models performed better on cross datasets evaluations. Thus, the CNNs models did better in the same train-to-test dataset evaluations, while the Transformers models performed better in the cross datasets evaluations. Overall, the XceptionNet models did well in both aspects.

Based on the cross datasets evaluations and overall models' performance, we observed closer relationships between the FF++ 2020 and Celeb-DF datasets, and the Google DFD and Celeb-DF datasets. However, even though the models trained on FF++ 2020 worked very well on the Celeb-DF test dataset, it did not for the other way round. This implied that the relationship was one-directional, and the FF++ 2020 dataset could be more diverse and possessed more super-set characteristics over the Celeb-DF dataset. This property may have been made prominent with the updated version of FF++ in 2020.

Models trained on the Deeper Forensics dataset fared poorly when tested on the other datasets. Some models trained using the DFDC dataset worked well when tested on the Deeper Forensics test dataset. Models trained on the DFDC dataset also performed very well on the other datasets, except for the FF++ 2020 test dataset. This implied that even as the DFDC dataset was quite diverse and comprehensive, it was still distinct from the FF++ 2020 dataset. Thus, the Deeper Forensics, DFDC and FF++ 2020 datasets remain important and warrant further investigations in future works along with newer and more sophisticated deepfakes datasets.

VI. CONCLUSIONS

In this work, we explored the second and third generation public deepfake datasets, and investigated how the evolving deep learning landscape will benefit deepfake detections. We designed and developed single model deepfake detectors based on eight different deep learning architectures (four CNNs and four Transformers) and conducted same train-to-test and cross datasets evaluations.

With our detection models, we achieved 88.74%, 99.53%, 97.68%, 99.73% and 92.02% accuracy and 99.95%, 100%, 99.88%, 99.99% and 97.61% AUC, in the detection of FF++ 2020, Google DFD, Celeb-DF, Deeper Forensics and DFDC deepfakes, respectively. We observed that CNNs models did better in same train-to-test dataset evaluations, and the Transformers models did better in cross datasets evaluations.

Based on further cross datasets evaluations and overall models' performance analysis, we showed the relationships among the FF++ 2020, Google DFD and Celeb-DF datasets. We also observed the strengths and uniqueness of the Deeper Forensics, DFDC and FF++ 2020 datasets and how they may continue to be important and warrant further investigations in future works along with newer and more sophisticated deepfakes datasets.